# Formulation and evaluation of ocean dynamics problems as optimization problems for quantum annealing machines

Takuro Matsuta[1*] and Ryo Furue[2]

[1] *Faculty of Environmental Earth Science, Hokkaido University, Hokkaido, Japan.*

[2] *JAMSTEC, Yokohama, Japan.*

[*]Corresponding author

Email: matsuta@ees.hokudai.ac.jp

# ABSTRACT

Recent advancements in quantum computing suggest the potential to revolutionize computational algorithms across various scientific domains including oceanography and atmospheric science. The field is still relatively young and quantum computation is so different from classical computation that suitable frameworks to represent oceanic and atmospheric dynamics are yet to be explored. Quantum annealing, one of the major paradigms, focuses on combinatorial optimization tasks. In this paper, we solve the classical Stommel problem by quantum annealing (QA) and simulated annealing (SA), a classical counterpart of quantum annealing. We cast the linear partial differential equation into an optimization problem by the least-squares method and discretize the cost function in two ways: finite difference and truncated basis expansion. In either case, SA successfully reproduces the expected solution when appropriate parameters are chosen, demonstrating that annealing has the potential. In contrast, QA using the D-Wave quantum annealing machine fails to obtain good solutions for some cases owing to hardware limitations; in particular, the highly limited connectivity graph of the machine limits the size of the solvable problems, at least under currently available algorithms. Either expanding the machine's connectivity graph or improving the graph-embedding algorithms would probably be necessary for quantum annealing machines to be usable for oceanic and atmospheric dynamics problems.

While this finding emphasizes the need for hardware improvements and enhancements in graph embedding algorithms for practical applications of quantum annealers, the results from simulated annealing suggest its potential to address practical geophysical dynamics problems. As quantum calculation continues to evolve, addressing these challenges may lead to transformative advancements in ocean and atmosphere modeling.

# 1. Introduction

The ocean and atmospheric circulations play a crucial role in shaping the climate system, necessitating accurate simulations of geophysical fluid dynamics for climate prediction. High-performance computing has successfully reproduced and predicted large-scale oceanic and atmospheric motions and climate changes [1–4]. However, scale separation does not exist for oceanic and atmospheric circulation and subgrid-scale parameterizations are not adequate for some problems. For example, the interaction between oceanic mesoscales, submesoscales, and internal waves, or interaction between coastal regions and the open ocean do not seem to allow for adequate subgrid-scale parameterizations. The only practical solution therefore seems to be to solve for all the relevant scales simultaneously. Until recently, Moore's law had allowed us to increase grid resolution as computers became exponentially faster year by year. That is unfortunately no longer the case. The alternative is to rewrite our simulation programs for specific hardware to improve performance. Despite the human-resource cost to do so, the gain will not be systematic and will sooner or later diminish as the code gets more and more complicated [5].

Quantum computing may be regarded as this approach but for radically different hardware using radically different programming methods [6–9]. There are two leading paradigms of quantum computation: One is gate quantum computing and the other is quantum annealing (QA). The former constructs a series of quantum operations on a set of quantum bits or "qubits", analogous to the logic circuit in classical computing [10]. The latter is designed to solve combinatorial optimization problems by controlling quantum fluctuations [11,12].

While quantum gate–based systems remain far from practical, QA machines provided by D-WAVE Inc. [13] have been put to practical use [14,15]. Notably, QA approaches have been tried in various industrial domains, including traffic flow optimization [16] and machine learning [17] (for a comprehensive review, refer to [15]). A potential advantage of QA is its potential speed in the future for large problems [6,18,19] although real annealing machines are currently not fast and often give solutions that include large error due to noise on hardware. In addition, QA machines possibly have higher power efficiency, i.e., computing power per Watt, than classical supercomputers [20].

To utilize QA in the context of atmosphere and ocean sciences, we cast geophysical fluid dynamics problems to equivalent optimization problems. As a first step, we choose simple linear partial differential equations that describe certain geophysical fluid dynamics problems. We test two methods. In one, we discretize the partial differential equation using finite differences. We can then solve the resultant linear equation using the standard least-squares method, which is an optimization problem. A previous study [21] tested this method but they had to use a large number of qubits to approximate real numbers, which makes it difficult to solve large problems. Indeed, their solutions included errors even for a small problem whose gridpoints are only 5. In this study, we show that an iteration procedure suggested by [22,23] dramatically improves accuracy. The other method we test is truncated spectral expansion [23], which approximates the solution by a truncated orthogonal basis series.

Our paper is structured as follows: We outline the quantum annealing method and its classical counterpart, "simulated annealing" (SA), in section 2. We also reformulate linear partial differential equations of geophysical fluid dynamics as an optimization problem by the least squares method. In section 3, we solve a simple wind-driven ocean circulation and compare the outcomes of QA with those of SA. We also compare the performance of the iteration method with the direct binary expansion with large spin numbers . In section 4, we extend the truncated spectral expansion algorithm to a very simple nonlinear problem in order to illustrate the method. Section 5 first summarizes the results and then compares our method with other potential quantum-computing methods.

# 2. Method

## 2.1 Background

In this study, we consider the linear partial differential equation of the form

$$L[f](\mathbf{x}) := \sum_k C_k(\mathbf{x})\partial^k f(\mathbf{x}) + B(\mathbf{x}) = 0, \tag{1}$$

where $\mathbf{x}$ is the position vector, $C_k$ is the variable coefficient, $\partial^k f$ is a vector containing all the partial derivatives of order $k$, and $B$ is the inhomogeneous term. Here, we assume a bounded rectangular ocean

$$X = \{\mathbf{x} = (x, y) \mid x \in [0,1], y \in [0,1]\}. \tag{2}$$

To obtain an approximate solution to (1), we minimize the following cost function [23,24]:

$$H = \|L[f]\|^2 := \int_0^1 \int_0^1 |L[f](x,y)|^2 \, dxdy, \tag{3}$$

where $\|\cdot\|$ (called $L^2$ norm) indicates the square root of squared-integral over the domain.

## 2.2 Discretization of the differential equation

To find a function $f$ that minimizes Equation (3) numerically, we discretize Equation (3) in two ways. One approach is to approximate Equation (3) with finite differencing, representing $f(x)$ with gridded values $w_{ij}$; and the other approach is to expand $f(x)$ with a truncated set of basis functions and express $H$ in terms of the expansion coefficients.

### 2.2.1 FINITE DIFFERENCING METHOD

If we discretize the domain $X$ into $N \times N$ grid points, as detailed in reference [25], Equation (1) can be discretized into a set of linear equations

$$A\mathbf{w} - \mathbf{v} = 0, \tag{4}$$

where $A = \{a_{ij}\}_{i,j\in\{1,2,...,N^2\}}$ is an $N^2 \times N^2$ matrix, $\mathbf{w} = \{w_i\}_{i\in\{1,2,...,N^2\}}$ is the solution we seek, and $\mathbf{v} = \{v_i\}_{i\in\{1,2,...,N^2\}}$ is the vector representing the inhomogeneous term. A least squares method forms the optimization problem

$$\mathbf{w} = \arg\min_{\widehat{\mathbf{w}}} \|A\widehat{\mathbf{w}} - \mathbf{v}\|^2 = \arg\min_{\widehat{\mathbf{w}}} H(\widehat{\mathbf{w}}) \tag{5}$$

where

$$H(\mathbf{w}) := \|A\mathbf{w} - \mathbf{v}\|^2 = \sum_{i,j=1}^{N^2} J_{ij} w_i w_j + \sum_{i=1}^{N^2} h_i w_i + \sum_{i=1}^{N^2} v_i^2, \tag{6}$$

$$J_{ij} = \sum_{k=1}^{N^2} a_{ki} a_{kj}, \tag{7}$$

and

$$h_i = \sum_{j=1}^{N^2} (-2 a_{ji} v_j). \tag{8}$$

Since the last term of Equation (6) is constant, it does not influence the optimization.

### 2.2.2 Truncated spectral expansion Algorithm

Following [23], we express the function $f$ as an expansion in terms of a truncated basis, $\{\phi_i\}_{i\in\{1,2,...,n_{basis}\}}$, as

$$f(\mathbf{x}) = \sum_{i=1}^{n_{basis}} w_i \phi_i(\mathbf{x}), \tag{9}$$

where $n_{basis}$ is the order of expansion and $w_i$ is the expansion coefficient. Then Equation (1) is written as

$$L[f](\mathbf{x}) \approx \sum_{i=1}^{n_{basis}} w_i G_i(\mathbf{x}) + B(\mathbf{x}), \tag{10}$$

where $G_i$ is defined as

$$G_i(\mathbf{x}) = \sum_k C_k(\mathbf{x}) \partial^k \phi_i(\mathbf{x}). \tag{11}$$

For the function $f$ to be an approximate solution to Equation (1), the cost function below should be minimized:

$$H = \int_0^1 \int_0^1 |L[f](\mathbf{x})|^2 dxdy \tag{12}$$

$$= \sum_{i,j=1}^{n_{basis}} J_{ij} w_i w_j + \sum_{i=1}^{n_{basis}} h_i w_i + \int_0^1 \int_0^1 |B(\mathbf{x})|^2 dxdy, \tag{13}$$

where

$$J_{i,j} = \int_0^1 \int_0^1 G_i(\mathbf{x}) G_j(\mathbf{x}) \, dxdy, \tag{14}$$

$$h_i = 2 \int_0^1 \int_0^1 B(\mathbf{x}) G_i(\mathbf{x}) \, dxdy. \tag{15}$$

The solution to the original problem is again converted to the minimization of a cost function. In the previous subsection (finite differencing), the conversion from $f(\mathbf{x})$ to $\mathbf{w}$ happened by finite differencing, and in the present subsection (truncated spectral expansion), the conversion happens by spectral expansion. A previous study [23] confirmed its efficacy in solving simple partial differential equations whose solutions can be well represented by a limited number of basis functions.

It should be noted that expression (9) is not an exact solution to the original differential equation; it is merely an approximate solution, and its accuracy inherently depends on where we truncate the spectral expansion. For this reason, the cost function does not approach zero. In contrast, the finite difference form (Equation (4)) has an exact solution, for which the finite-difference cost function vanishes.

## 2.3 Transformation of cost function into an Ising model

We describe basic principles of the Ising model and its connection to the minimization problem discussed in Section 2.2. The Ising model was originally proposed as a model to understand ferromagnetism. It consists of $N$ spins located at lattice sites. Each spin can exist in one of two states: "upward" or "downward." The Ising Hamiltonian, representing the energy of the system, is given by

$$H_0 = -\sum_{i,j \in V} J_{ij} \sigma_i \sigma_j - \sum_{i \in V} h_i \sigma_i, \tag{16}$$

where $V = \{1, 2, \cdots N\}$ is the set of the spin sites, $\sigma_i$ is a *spin variable*, a type of binary variable that can take 1 or −1, $J_{ij}$ is an $N \times N$ matrix representing the interaction between the *i*-th and *j*-th spins, and $h_i$ is the external field that acts on the *i*-th spin. The state $\boldsymbol{\sigma} = \{\sigma_1, \ldots, \sigma_N\}$ that minimizes the Hamiltonian is known as the *ground state*.

The form of cost functions introduced in Section 2.2 (Equations (6) and (13)) is identical to that of the Ising Hamiltonian (Equation (16)) except for whether the variables are real or binary and except for the constant term. To transform the cost functions to the Ising

Hamiltonian, the real-valued variables, **w** of Equations (6) and (13) need to be approximated by the spin variable. We first expand **w** in terms of spin variables $\sigma$ as

$$w_i = c_i + s_i \sum_{\alpha=0}^{n_{spin}} \frac{\sigma_i^{(\alpha)}}{2^\alpha}, \tag{17}$$

where $\sigma_i^{(\alpha)} \in \{-1,1\}$ is the spin variable and $n_{spin}$ is the number of spins per variable, $s_i$ is a *scale factor* and $c_i$ is a *control parameter*. Substituting the binary expansion ( (17)) to the cost function (See Equations (6) or (13)), we obtain the Ising Hamiltonian. If the scale factor and control parameter are appropriately chosen, the spin variables of the ground state give the global minimum of the cost function up to a precision of $2^{-n_{spin}}$.

## 2.4 Annealing method

The annealing is a general algorithm that searches for the ground state of an Ising model. We introduce two types of annealing: simulated annealing (SA) and quantum annealing (QA).

### 2.4.1 SIMULATED ANNEALING

SA is a probabilistic approach. We investigate the characteristics of the SA approach as the classical counterpart of QA in this study. (SA itself has been used for oceanographical optimization problems outside the context of quantum annealing [26–28])

During each iteration of SA, a single spin, chosen at random, is inverted. If this change results in a reduction of the Hamiltonian, the new spin configuration is accepted. Conversely, if the Hamiltonian increases, the new configuration may still be accepted, albeit with a less-than-one probability. The Metropolis algorithm, as shown in [29], is frequently employed to determine this probability of acceptance:

$$P(\Delta E, T) = \begin{cases} 1, & \Delta E \leq 0 \\ \exp\left(-\dfrac{\Delta E}{k_B T}\right), & \Delta E > 0 \end{cases}, \tag{18}$$

where $T$ is the "temperature", $k_B$ is the Boltzmann constant, and $\Delta E$ is the energy difference of the new configuration from that of the old configuration. It is noteworthy that the second row is the Boltzmann distribution. In this framework, SA conducts a ground state search through simulated thermal fluctuations. If the system is trapped in a local minimum, these fluctuations facilitate an escape. This allows the continued pursuit of the global minimum. This "hill-climbing" feature can be a significant advantage over gradient descent algorithms [29].

We repeat the Metropolis algorithm by decreasing the temperature in each step, which is equivalent to the reduction of the probability of accepting worse solutions. If the schedule of the temperature decreasing is sufficiently slow in such a way that

$$T(t) \geq \frac{aN}{\log(\alpha t + 2)}, \tag{19}$$

the SA gives the true solution [29,30]. Here, $a$ and $\alpha$ are some constants. This condition shows that the time for $T(t)$ to reach a sufficiently small value $\epsilon$ is

$$t \sim \exp\left(\frac{N}{\epsilon}\right), \tag{20}$$

indicating that the temperature schedule should be exponentially slower for a larger number of spins. It is noteworthy that the condition is not an optimal estimate but just the worst-case estimate.

## 2.4.2. QUANTUM ANNEALING

The QA approach is also an algorithm that searches for the ground state of the Ising model. While the SA utilizes simulated thermal fluctuations, quantum fluctuations are responsible for the search in QA [12]. We provide a brief introduction to QA in this section, and Appendix provides an intuitive introduction to QA for readers unfamiliar with the quantum mechanism.

We define a time-dependent Hamiltonian of the system as follows:

$$\widehat{H}(t) = \widehat{H}_0 + \Gamma(t)\widehat{H}_i, \tag{21}$$

where $\widehat{H}_0$ is the quantum Ising Hamiltonian, $\widehat{H}_i$ is an initial Hamiltonian, and $\Gamma(t) \geq 0$ is a monotonically decreasing function of time representing the annealing "schedule" [11,12,14]. To differentiate it from the classical Hamiltonian, the quantum version is denoted with a hat symbol. In the context of QA, a "transverse field" is often selected as the initial Hamiltonian since its ground state is known and easily prepared. This field drives the quantum fluctuation and hence corresponds to the thermal fluctuation of SA. Initially, the qubits are aligned with the ground state of this initial Hamiltonian. The amplitude of the transverse field, denoted by $\Gamma(t)$, is then gradually reduced from a substantially large value to zero. If the state consistently stays in the ground state of $\widehat{H}(t)$ at all times, it will naturally converge to the ground state of $\widehat{H}_0$, which is the solution we seek.

The *adiabatic theorem* [31] ensures this expectation, provided changes in $\Gamma(t)$, i.e., $\frac{d\widehat{H}}{dt}$, is sufficiently slow. According to [30,32], a sufficient condition of convergence of QA is

$$\Gamma(t) \geq a(\delta\, t + c)^{-1/(2N+1)}, \tag{22}$$

where $a$ and $c$ are constants, $\delta$ is a small parameter, and $N$ is the number of qubits. This condition shows that the time for $\Gamma(t)$ to reach a sufficiently small value $\epsilon$ is

$$t \sim \exp(N|\log \epsilon|). \tag{23}$$

This condition is not an optimal estimate but a worst-case one as in the case of SA. A comparison of Equations (20) and (23) shows that both SA and QA take exponential time to converge at worst, but the QA approach is better than the SA approach since $|\log \epsilon| \ll \epsilon^{-1}$ for $0 \leq \epsilon \ll 1$.

## 2.5 Implementation of annealing for practical problems

To obtain a solution, which is a set of real numbers for us, on an annealing machine, which uses binary variables (spins), there are two methods. One is to express each real number as a series of binary variables (Equation (17)) [21]. Although straightforward, this approach reduces the likelihood of attaining the global minimum due to increased spin numbers. In addition, the required annealing time increases exponentially with $N$ at worst (Equations (20) and (23)).

Alternatively, we use an iteration procedure [22,23] with a fixed small $N$. In each iteration $I$, which we call an *epoch*, the hyperparameters of Equation (17) are updated as follows:

$$c_i^{(I)} = w_i^{(I-1)}, \tag{24}$$

$$s_i^{(I)} = S s_i^{(I-1)}, \tag{25}$$

where $S$ is a scale factor to "zoom in" in each epoch. Simply put, we initially obtain a very crude approximate solution with $w_i = \pm 1$. Next, we look for a better approximation around this initial solution, narrowing the range of the search by a factor of $1/S$ and rewriting $H$ in terms of the new $w_i$. We repeat this "zooming extension" until we arrive at a desired precision. A notable potential drawback is that if incorrect values are obtained for $w_i$ in an earlier iteration, they cannot be corrected in the subsequent iterations unless the scale factor $S$ is close to 1 but the convergence is slower then. We will discuss the impact of the parameter *S* on accuracy in Section 3.

## 2.6 Programming an annealing machine

We implement our SA code using *Fixstars Amplify SDK*, a Python library for formulating combinational optimization problems [33]. The SDK provides data structures for SA and functions to operate on them. Spins of SA are fully connected; that is, $J_{ij}$ can be nonzero for any pair of (*i,j*). The annealing schedule is automatically optimized by the SDK in this study.

For QA, we use the D-Wave annealer *Advantage_system4.1*. Because of physical noise, identical annealing runs give different results. For this reason, we run QA 500 times per iteration and choose the best solution, that is, the one that gives the smallest Hamiltonian value. Unlike SA, the D-wave annealer has limited connectivity, that is, $J_{ij}$ has to be zero for a large number of (*i,j*) pairs; in the Isign-model terminology, not all pairs of spins can interact. Only "neighbors" can interact and "long-range" interactions do not exist. This is a serious problem because, for example, spectral expansion naturally includes long-range interactions.

To go around this limitation, the Hamiltonian is transformed by embedding the connectivity graph of $J_{ij}$ in the graph of hardware [34] (D-Wave forms what is called a Pegasus graph.) To illustrate graph embedding, we consider a simple virtual machine whose spin connectivity graph is represented by **Figure 1**(a). We try to represent the Hamiltonian

$$H = \sigma_0\sigma_1 + 2\sigma_1\sigma_2 - 3\sigma_0\sigma_2 + 4\sigma_2\sigma_3 - 5\sigma_3\sigma_4 - 6\sigma_2\sigma_4. \tag{26}$$

on this virtual machine. The connectivity of the Hamiltonian is shown in **Figure 1**(b). To represent this connectivity in the graph of the machine (**Figure 1**a), two qubits of the machine are assigned to represent the single spin $\sigma_2$ of the Hamiltonian as shown in **Figure 1**(c). To ensure that the two qubits have the same value, a penalty term is added to the Hamiltonian. The penalty term must be large enough to ensure that the two qubits are the same but small enough not to dominate the optimization and lead to a solution far from the ground state of the original Hamiltonian. This trade-off can sometimes be difficult [35]. For this and other reasons, graph embedding by itself is a subject of active research. In this paper, we use a library in "Fixstars Amplify SDK" for this transformation of the Hamitonian. **Figure 2** summarizes the procedures of SA and QA.

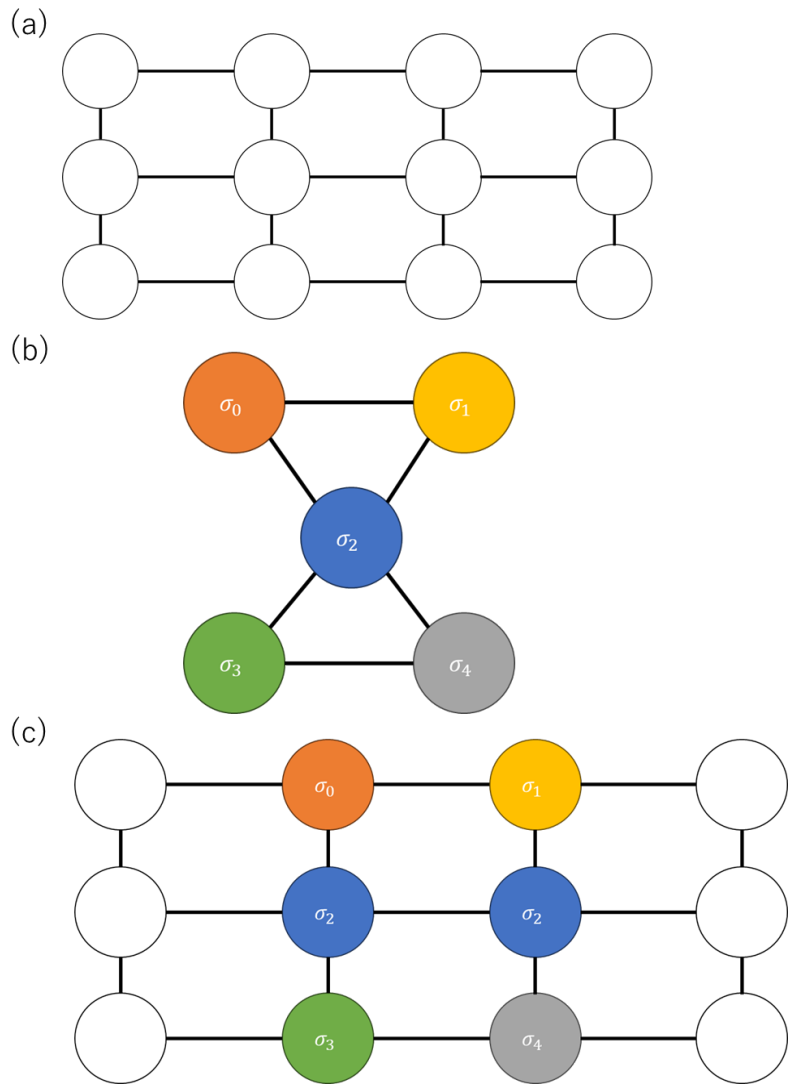

**Figure 1. Schematics for graph embedding.** Schematics showing (a) the spin configuration of a virtual annealing machine, (b) the spin configuration of the Hamiltonian (Equation (26)), and (c) how the Hamiltonian is embedded in the machine's connectivity graph.

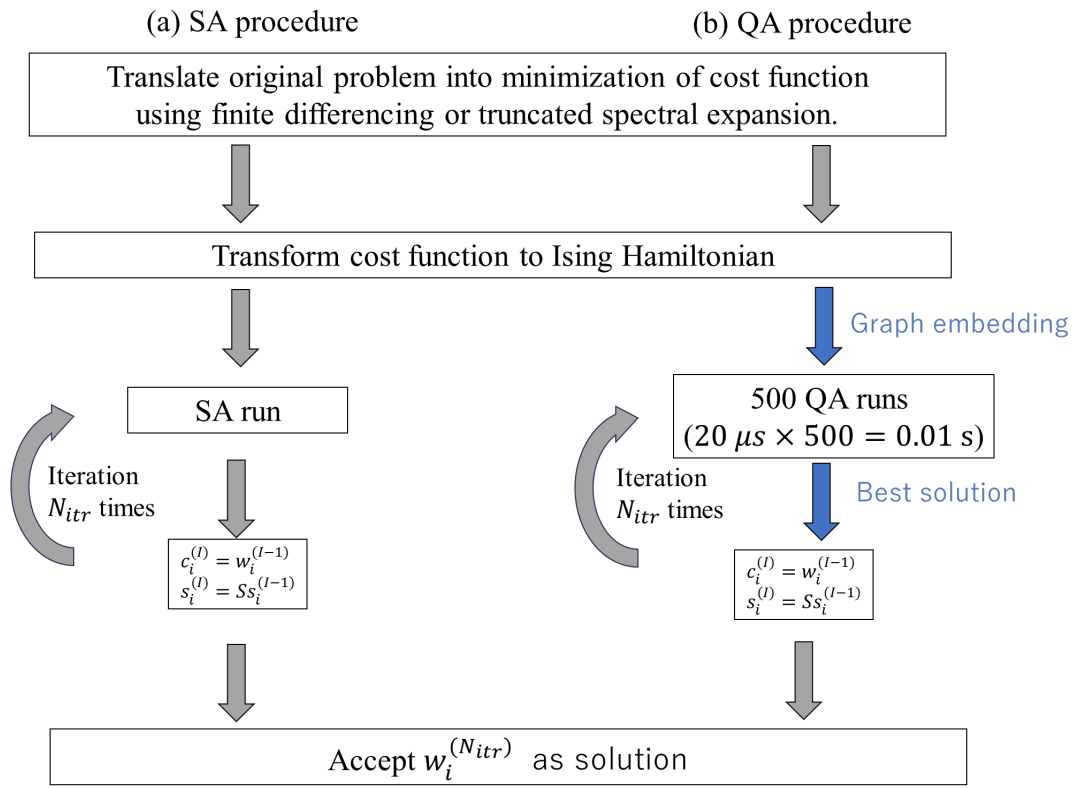

**Figure 2. Schematics of SA and QA procedure.** Schematic of procedure of (a) SA and (b) QA approaches.

# 3. The Stommel problem

In this section, we solve the "Stommel problem", comparing finite difference with truncated spectral expansion, and SA with QA , and iteration with spin series by large spin number. According to [36,37], the streamfunction of a wind-driven gyre forced by zonal wind stress is described by the partial differential equation

$$\frac{\partial \psi}{\partial x} + \epsilon \left( \frac{\partial^2 \psi}{\partial x^2} + \frac{\partial^2 \psi}{\partial y^2} \right) + \frac{\partial \tau_x}{\partial y} = 0, \qquad (32)$$

where $\psi$ is the streamfunction, $x \in [0,1]$ and $y \in [0,1]$ are the zonal and meridional position, $\tau_x = -\cos(\pi y)$ is the zonal wind stress, and $\epsilon$ is a small parameter associated with friction. We assume that $\psi = 0$ at the lateral boundaries. Here, all variables are nondimensionalized and we set $\epsilon = 0.1$ as an example.

## 3.1. Finite-difference solution using SA.

We discretize the domain into an 11×11 grid. A standard finite differencing transforms Equation (32) into a set of 121 linear equations. The boundary conditions are already encoded in this set. The set is solved using the least squares method (section 2.2.1). We first solve the least-squares problem using standard linear algebra. We refer to this solution as "true" and compare it with solutions from the annealing methods. The linear algebra calculation implemented using LAPACK takes 0.03 s on the lead author's laptop PC, which corresponds to one iteration time of QA (see section 2.6) on the actual D-wave machine. At this point in time, annealing is not faster than classical linear algebra on actual machines for our problem at hand.

To investigate how the annealing solution depends on the hyperparameters, we conduct SA experiments with $(S, n_{spin}) = (0.5, 2)$, $(S, n_{spin}) = (0.5, 3)$, $(S, n_{spin}) = (0.8, 2)$, and $(S, n_{spin}) = (0.8, 3)$. **Figure 3** and **Figure 4** plot the streamfunction and the cost function. With $(S, n_{spin}) = (0.5, 2)$, the SA solution is significantly different from the true one (**Figure 3**a) even though it is well converged (**Figure 4**a). Note that the cost function converges to a much higher value (**Figure 4**a) than for the better solutions (**Figure 4**b-d), which indicates that this one has converged to an incorrect solution.

Increasing the number of spins improves accuracy. When $n_{spin} = 3$, the cost function becomes much smaller (**Figure 4**b) and the solution is improved (**Figure 3**b). Setting $S = 0.8$, however, gives an even better solution even with $n_{spin} = 2$ (**Figure 3**c). The convergence is much slower (**Figure 4**c) and the accuracy still keeps improving after 30 epochs. Increasing the number of spins from 2 to 3 again does not give a significant improvement (**Figure 4**c and **Figure 4**d).

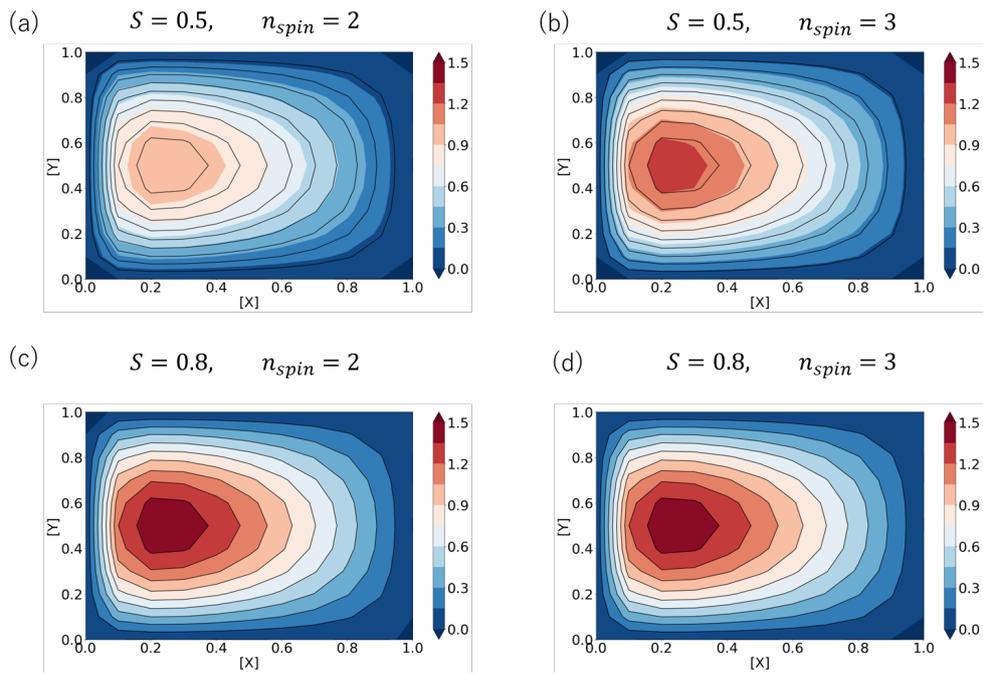

**Figure 3. Streamfunction calculated by SA with different hyperparameters.** Streamfunction from SA (shading) and the "true" solution (black contour lines). The hyperparameters are (a) $(S, n_{spin}) = (0.5, 2)$, (b) $(S, n_{spin}) = (0.5, 3)$, (c) $(S, n_{spin}) = (0.8, 2)$, and (d) $(S, n_{spin}) = (0.8, 3)$.

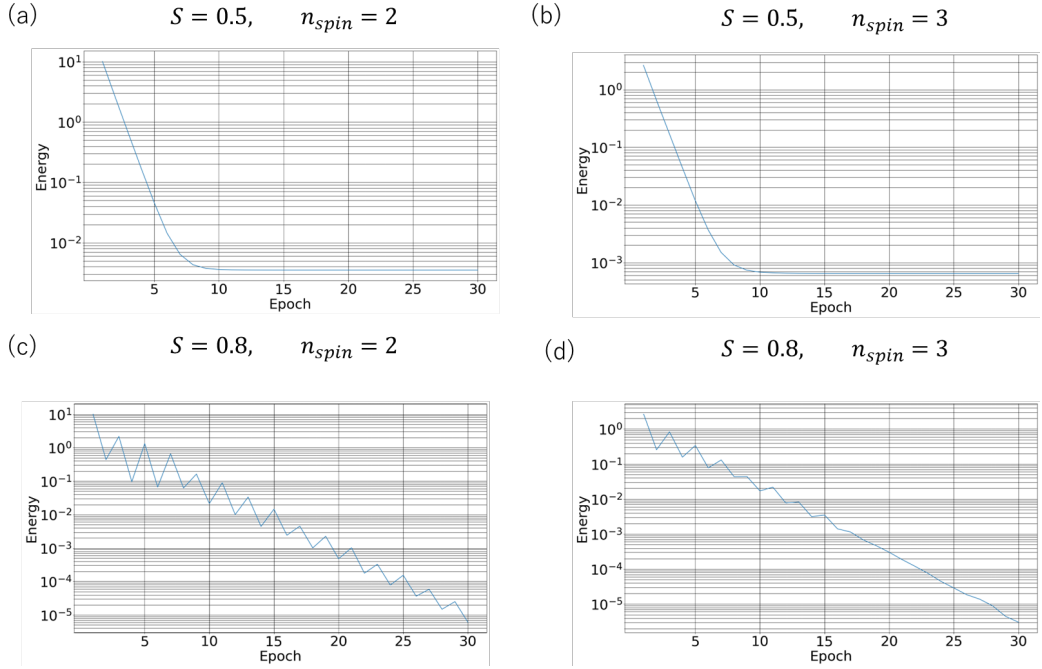

**Figure 4. Cost function calculated by SA with different hyperparameters.** Cost function (energy) versus epoch for the same four experiments as in **Figure 3**.

## 3.2. QA solution to the finite-difference form

We solve the same problem using the D-Wave annealer. We only test the $(S, n_{spin}) = (0.8, 2)$ case. **Figure 5**(a) shows that the QA approach fails to solve the Stommel problem. Accordingly, the cost function remains larger than 1.0 (**Figure 5**(b)). There are three possible reasons for this failure. One is that the required annealing time is much larger than those we have tested because of the increased number of qubits needed to embed the graph (Section 2.6). Another is that the penalty terms introduced for the embedding are so large as to dominate the minimization, which is leading to incorrect solutions (Section 2.6). The other is that solutions from the machine always include error due to noise on the hardware [19].

We next reduce the resolution to a 5×5 grid with $\epsilon = 0.25$. As shown in **Figure 6**(a), the accuracy is dramatically improved and the value of the cost function is below $10^{-2}$.

In theory, all optimization problems of the form (21) can be solved if the annealing time is sufficiently increased (Section 2.4.2) but obviously the graph embedding would lower the performance. The above results therefore suggest that better graph embedding algorithms would improve performance even for large problems. Equation (32) is a two-dimensional Poisson equation and the matrix of its finite-difference version is a very sparse matrix whose graph can be significantly different from the machine's native graph, necessitating good embedding algorithms. Given that the Poisson equation arises in important oceanographic and meteorological techniques (such as calculating wave energy flux [38] or Helmholtz decomposition for atmospheric circulations [39]), it would be worthwhile to implement a multi-dimensional Poisson solver using QA.

(a)

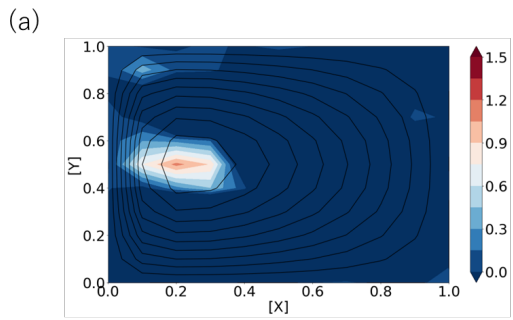

(b)

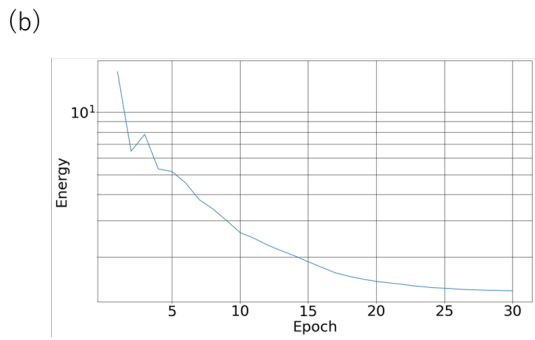

**Figure 5. QA solutions for the Stommel problem.** (a) Streamfunction from QA (shading) and the "true" solution (black contour lines).   (b) Cost function (energy) versus epoch.

(a)

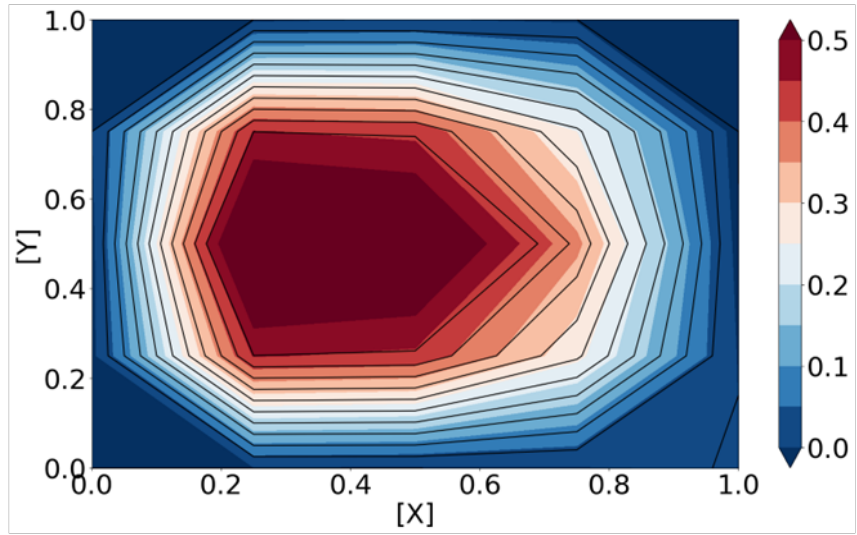

(b)

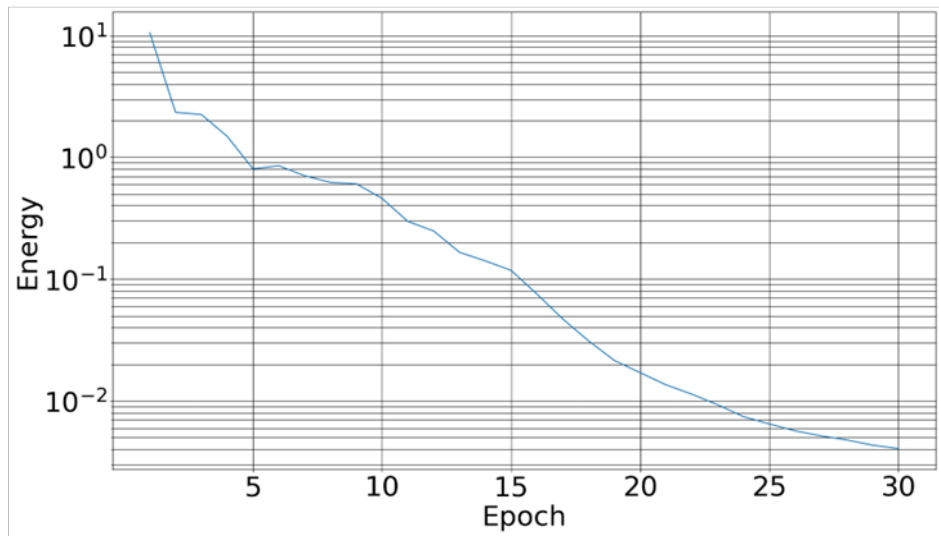

**Figure 6. QA solutions for the coarse Stommel problem.** Same as **Figure 5** but when the model domain is discretized into $5 \times 5$ grid and $\epsilon = 0.25$. Hyperparameters are set to $(S, n_{spin}) = (0.8, 3)$.

## 3.3. QA and SA solutions for truncated spectral expansion

We apply the truncated spectral expansion approach to the Stommel problem. Given that the streamfunction must vanish at the lateral boundaries, we approximate it using a low-order truncated Fourier sine expansion

$$\phi(x,y) = \sum_{n,m} w_{nm} \sin(\pi n x) \sin(\pi m y). \tag{33}$$

Here, we assume that $n \in \{1,2,\ldots,n_x\}$ and $m \in \{1,2\}$. We use only 2 modes in the $y$ direction because we know that the solution will be very smooth in the $y$ direction, while we set $n_x = 10$. Since the SA method stably returns the global optimization as seen in previous sections, we compare the QA results with the SA results as a "true" solution.

**Figure 7** compares the results from the SA and QA approaches. When $n_x = 10$, the SA (black contours) and QA solutions (color shading) are almost identical. The convergence of QA (**Figure 7**(b)) is also similar to that of SA (not shown).

Although the QA approach returns the solutions with the level of accuracy as SA, the amplitude of the gyre is smaller than that obtained from the finite-difference form (**Figure 3**) due to the truncation error. Theoretically, this error can be reduced by an increase in the expansion degree. However, the performance dramatically decreases due to the graph embedding and noise on the hardware. Indeed, the QA does not return a reasonable solution with a gyre structure and the cost function converges to $\mathcal{O}(10^2)$ even after 30 iterations when $n_x = 20$ (not shown). Hence the truncated spectral expansion approach can give a crude approximation, but improving the accuracy of solutions is challenging.

(a)

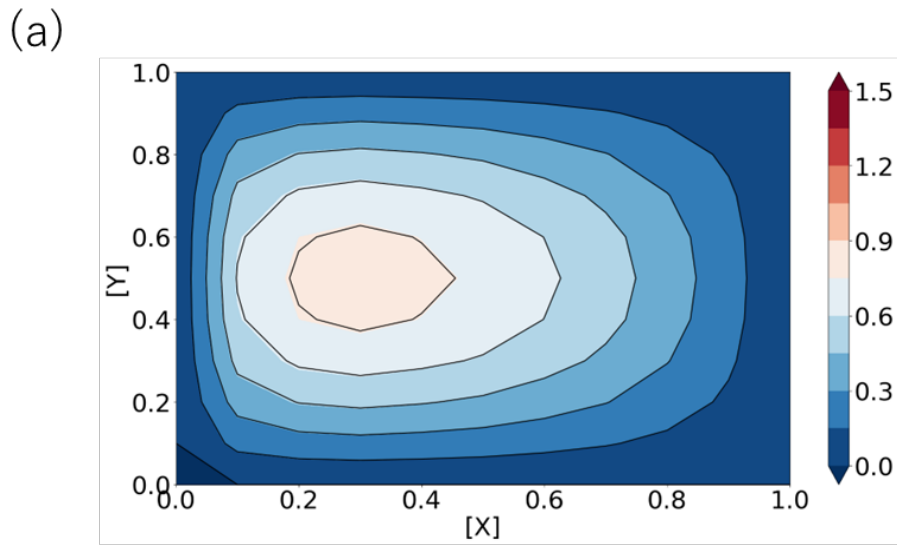

(b)

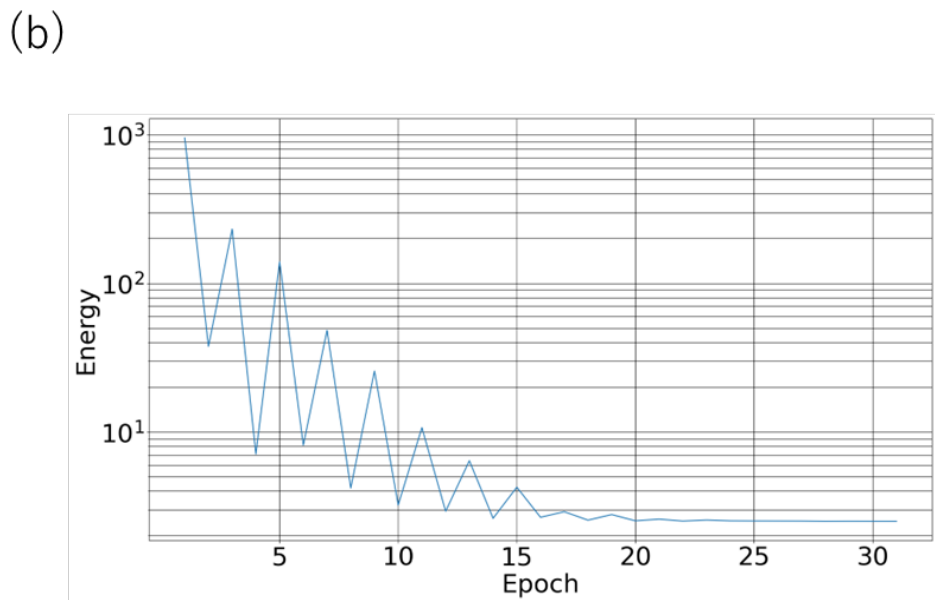

**Figure 7. QA and SA solutions by truncated spectral expansion.** (a) Streamfunction from QA (shading) and SA (black contour lines). (b) Cost function versus epoch of QA.

## 3.4. Iteration procedure versus large spin number

So far, we have used iteration to obtain approximations to real numbers. In this section, we compare this method with the one using a series of spins to express each real number (section 2.5). We discretize the domain into an 5×5 grid and set $\epsilon = 0.25$ for the Stommel model with finite difference. As an example, we test a case with $n_{spin} = 5$ and another with $n_{spin} = 8$.

The solution from SA is good (**Figure 8**) and the cost function is $\mathcal{O}(10^{-1})$ and $\mathcal{O}(10^{-2})$. In contrast, we needed 10 iterations (epochs) with $n_{spin} = 2$, to reduce the cost function to the same level (not shown), indicating that increasing the number of spins is more efficient than the iteration procedure.

By contrast, the QA method does not produce a good solution without iteration (**Figure 9**) even though the annealing time is set to $1000 \, \mu s$. The cost function remains $\mathcal{O}(1.0)$ in this case. The failure can be attributed to noise as well as to the graph embedding (see also Section 3.2). According to [19], the D-Wave annealer possibly performs worse than the classical computer when the long annealing time is $\mathcal{O}(100\mu s)$ or longer. A larger number of spins generally requires more annealing time, but noise on the hardware reduces the likelihood of attaining the global minimum.

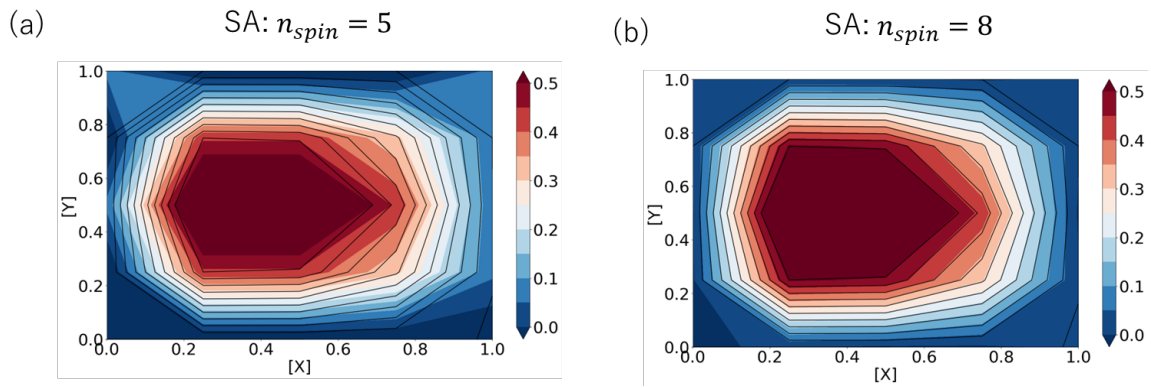

**Figure 8. SA solutions using a series of spins.** Streamfunction from SA (shading) and the "true" solution (black contour lines). The hyperparameters are (a) $n_{spin} = 5$ and (b) $n_{spin} = 8$.

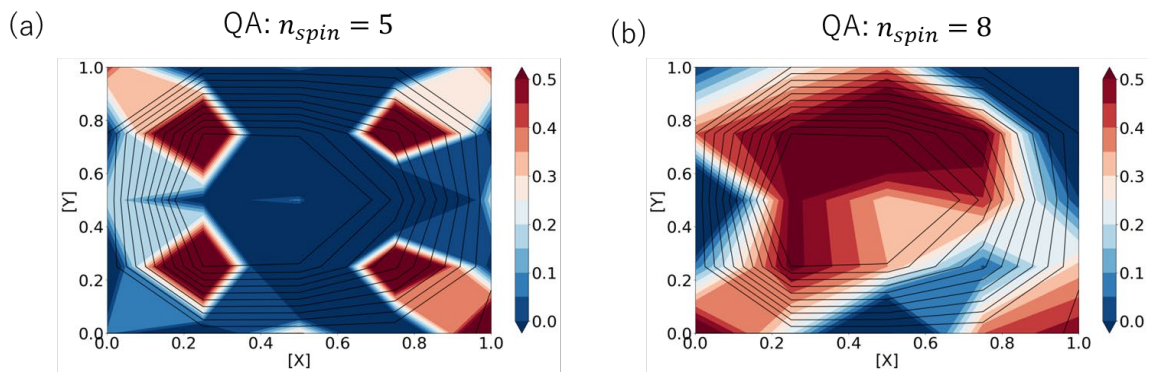

**Figure 9. QA solutions using a series of spins.** Same as **Figure 8** but from QA.

# 4. Extension of truncated spectral expansion to nonlinear differential equations

As we have seen in the preceding sections, a linear equation can be transformed into a minimization problem of a quadratic form like Equation (6) and renders itself amenable to annealing. What about nonlinear problems as more realistic ocean and atmospheric problems are? At this moment, we cannot handle realistic nonlinear problems. Instead, we use an abstract simple nonlinear differential equation to illustrate how the truncated spectral expansion approach can be applied to nonlinear differential equations. We consider a simple nonlinear ordinary differential equation defined in $x \in [0,1]$

$$\left(\frac{dy}{dx}\right)^2 - 4x^2 = 0, \tag{34}$$

with the boundary conditions that $y(0) = 1$ and $y(1) = 2$. The true solution of this equation is $y_{\text{true}} = 1 + x^2$. We expand the solution in terms of basis (as in Equation (9)). Here, we employ polynomials $\phi_m = x^m$ as the basis and expand the solution as

$$y = \sum_{m=0}^{n_{basis}-1} w_m \phi_m. \tag{35}$$

We use Equation (17) to express $w_m$ in terms of binary variables. As before (section 2.2.2), we seek an approximate solution that minimizes $H = \|(dy/dx)^2 - 4x^2\|^2$. After a straightforward calculation, we find

$$H = \sum_{i,j,p,q} \tilde{J}_{ijpq} w_i w_j w_p w_q + \sum_{i,j} J_{ij} w_i w_j + \int_0^1 16x^4 dx, \tag{36}$$

where

$$\tilde{J}_{ijpq} = \int_0^1 \phi'_i \phi'_j \phi'_p \phi'_q dx, \quad (37)$$

and

$$J_{ij} = \int_0^1 (-8x^2) \phi'_i \phi'_j dx. \quad (38)$$

represent interactions. Here, $\phi'_i$ indicates the derivative of $\phi_i$. Since the first term of the cost function involves interactions of four spins, we replace higher order terms by quadratic polynomials using Ishikawa's algorithm [40]. The degree reduction algorithm is implemented in *Fixstars Amplify SDK*.

The nonlinear equation (34) is solved using both the SA and QA approaches. We set $n_{basis} = 4$, and $(S, n_{spin}) = (0.8, 3)$. Note that the exact solution is $(w_0, w_1, w_2, w_4) = (1, 0, 1, 0)$ because $y_{true} = 1 + x^2$. **Table 1** shows that both SA and QA reproduce the true solution, but the accuracy of QA is lower. The value of the cost function is below $10^{-4}$ in QA and $10^{-7}$ in SA after 30 iterations.

**Table 1. SA and QA solutions for the simple nonlinear differential equation.** Each Value indicates the expansion coefficient obtained from the SA and QA approaches.

|    | $w_0$ | $w_1$ | $w_2$ | $w_3$ |
|----|-------|-------|-------|-------|
| SA | 1.00  | $-4.04 \times 10^{-4}$ | 1.00 | $-4.01 \times 10^{-4}$ |
| QA | 1.00  | $2.10 \times 10^{-2}$ | 0.964 | $1.85 \times 10^{-2}$ |

## 5. Summary

In this study, we have explored potential feasibility and potential problems of the annealing approach for atmospheric and oceanic problems. We consider linear equations, $L[f] = 0$, and obtain (approximate) solutions to them that minimize the cost function $H = ||L[f]||^2$. We discretize the cost function in two ways. The first is standard finite-difference and the other is truncated spectral expansion. After the discretization, the cost function takes the form of a quadratic form (Equations (6) & (13) ). The solution we seek is a set of real numbers $\{w_1, w_2, \dots\}$. Quantum annealing, however, operates on spin variables $\{\sigma_1, \sigma_2, \dots\}$, which take values of either 1 or $-1$. There are two methods to solve for $w$'s using $\sigma$'s. One is to approximate each $w_i$ as a series of multiple σ's (Equation (17)). The other is to search for the solution $w_i$ by iteration using a small number of $\sigma_i$ for each $w_i$ at each iteration (section 2.5). The latter method uses fewer spins at the cost of iterations. Here lies a trade-off point: the time it takes for an annealing machine to reach a good solution grows rapidly with the number of spin variables the problem uses. Using either method, the cost function is rewritten into a quadratic form in terms of spin variables. This form is often called "Ising Hamiltonian". As a quadratic form, an Ising Hamitonian is written as $H = \sum_{i,j} J_{ij} \sigma_i \sigma_j + \sum_i h_i \sigma_i$. Matrix $J_{ij}$ represents connectivity between spins $i$ and $j$; where $J_{ij} = 0$, there is no connection between those spins. As such, the non-zero parts of $J$ can be thought of forming a connectivity "graph". Vector $h_i$ represents the external field on each spin.

We compared quantum annealing (QA) with simulated annealing (SA), which also solves the minimization problem of an Ising Hamiltonian using a simulated and idealized physical process. It also includes an adjustable parameter that must be increased to obtain a good solution as the number of variables increases, which slows down the simulation. Apart from the fact that QA is carried out on a real machine whereas SA is a software simulation, the most significant difference between QA and SA to our problems is that real machines can handle only a very limited connectivity that is hardwired to them. For this reason, we often have to transform the actual graph *J* of the problem into a graph that can be embedded in the machine's native graph, which introduces extra spin variables and increases the potential for incorrect solutions (Section 2.6).

We applied QA and SA to the Stommel problem as an illustrative example of simple oceanographic problems. With the finite differential method, the SA approach gave accurate solutions. We furthermore demonstrated that appropriate hyperparameters enhance the accuracy more efficiently than merely increasing the number of spins. QA, on the other hand, gave incorrect solutions when the special resolution of finite differencing was high (corresponding to $n_{spin} = \sim 10^2$), while QA gave a relatively accurate solution when the resolution was low (corresponding to $n_{spin} = \sim 10^1$). (Of course, the solution is very crude because of the extremely low spatial resolution.)

QA and SA with truncated spectral expansion showed similar performance, although the accuracy of solutions was lower than with the finite differential method. One would expect that the accuracy can be improved by increasing the order of spectral expansion. That is, however, not the case: the QA converges to wrong solutions. Since spectral modes extensively interact with each other, the connectivity graph is possibly much denser than with finite difference and graph embedding introduces much more extra spin variables, rendering this method less practical.

Furthermore, we compared the iteration method with the non-iterative method that uses a larger number of spins. QA failed to reproduce the solution even for a coarse problem. The larger number of spins must have been the ultimate cause of the failure. The larger number of spins requires more annealing time (section 2.4); in addition, it further complicates the graph embedding, further increasing the number of spins (section 2.6). However, SA reproduced the true solution by the non-iterative method, suggesting that future improvements in QA hardware will make QA more practical. If thermal noise and other imperfections are suppressed and better embedding algorithms are developed, QA will be potentially faster than SA for some problems [6,18,19].

In this paper, we have focused on QA. The other major approach, quantum-gate computing, also has the potential to solve oceanic and atmospheric problems. Recent studies have successfully solved nonlinear fluid-dynamical problems on the basis of quantum-gate algorithms [41–43]. Tensor networks can also be used to solve nonlinear fluid-dynamical problems [44,45]. Although the tensor network is a classical algorithm, it can be implemented on quantum-gate machines [44,45]. There are still many challenges before weather and climate simulations become feasible on quantum computers [9]. It is an exciting time.

# Appendix

We provide an intuitive introduction to quantum annealing (QA) for oceanographers and meteorologists who are not familiar with quantum mechanics or statistical physics. We consider the minimization problem of

$$H_0 = \sigma_1\sigma_2 + \sigma_1 - \sigma_2, \tag{S1}$$

where $\sigma_1$ and $\sigma_2$ are "spin variables" that take +1 (upspin) or −1 (downspin). The first term originates from an interaction between qubits 1 and 2. The second and third terms are an external field acting on qubits 1 and 2, respectively. In terms of the notation of the Ising Hamiltonian (Equation (16) in the main text), $J_{12} = J_{21} = -1/2$, $J_{11} = J_{22} = 0$, $h_1 = -1$, and $h_2 = 1$ in this case. The minimum value of this Hamiltonian is −3, which occurs when $(\sigma_1, \sigma_2) = (-1, 1)$. In general, the state $(\sigma_1, \sigma_2, \dots)$ in which the Hamiltonian takes its global minimum is called the "ground state". Figure S1 is a schematic representation of this Ising model.

To find the ground state by the QA procedure, we first add quantum fluctuation by "transverse field". This effect is represented by $\Gamma(t)H_i$ in Equation (21) in the main text. The time-dependent constant $\Gamma(t)$ corresponds to the strength of the quantum fluctuation [30]. Initially, the quantum fluctuation is strong, and hence the ground state $\psi$ of the modified Hamiltonian $H_0 + \Gamma H_i$ is represented as the superposition of all possible spin configurations,

$$\psi(t=0) = \frac{1}{2}(\psi_{\uparrow\uparrow} + \psi_{\uparrow\downarrow} + \psi_{\downarrow\uparrow} + \psi_{\downarrow\downarrow}). \tag{S2}$$

Here $\psi_{\uparrow\downarrow}$, for example, indicates the spin configuration where qubit 1 is upward and qubit 2 is downward. Equation (S2) indicates that each possible spin configuration is observed with a probability of $1/4$ initially (Figure S2). We then decrease $\Gamma$ to zero sufficiently slowly. As a result, the quantum superposition diminishes and converges to the ground state, $\psi_{\downarrow\uparrow}$, (Figure S2), of the original Hamiltonian, $H_0$. The adiabatic theorem [31] guarantees the convergence to the ground state of $H_0$ under the condition of Equation (22) in the main text.

In summary, QA searches for the global minimum of the cost function using the quantum superposition of $2^N$ possible spin configurations simultaneously, where $N$ is the number of qubits. The worst-case time QA takes is $\sim \exp(N|\log\epsilon|)$ according to Equation (23) in the main text, but if QA takes only a polynomial time, that is, $O(N^p)$, QA will be faster for

sufficiently large $N$ than on classical machines, which requires $\mathcal{O}(2^N)$ calculations in general. Even in the worst cases, QA may still be faster if $\epsilon$ is sufficiently small.

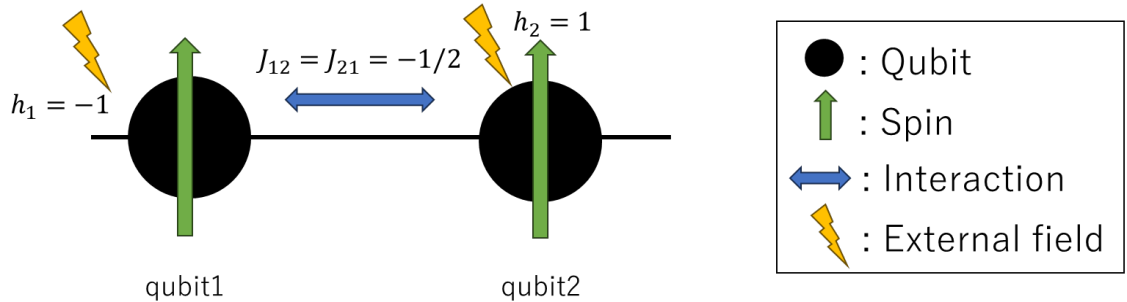

**Figure S1. Schematic of the Ising model associated with Equation (S1).** The figure shows the Ising model when both spins are upward. Green arrows indicate the spins.

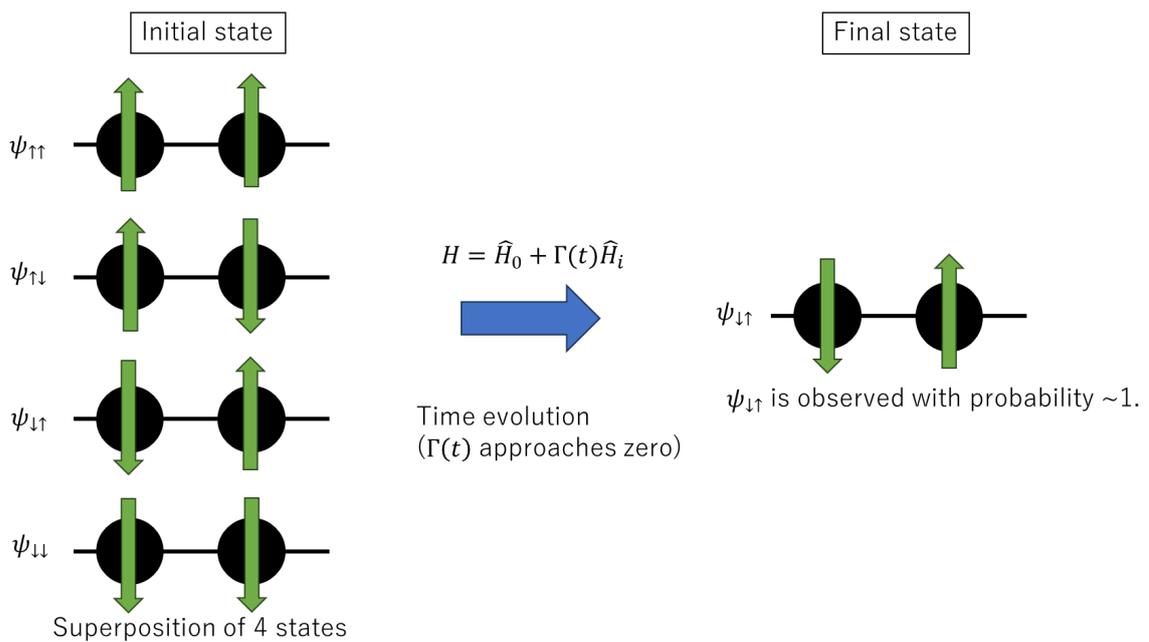

**Figure S2. Schematic of the QA procedure for the Ising model of Figure S1.** Left pannel indicates the initial state of the QA and the right panel indicates the state after the QA.